\documentclass{sigchi-ext}
\usepackage[T1]{fontenc}
\usepackage{subcaption}
\usepackage{textcomp}

\usepackage[scaled=.92]{helvet} 
\usepackage{graphicx} 
\usepackage{balance}  
\usepackage{booktabs} 
\usepackage{ccicons}  
\usepackage{ragged2e} 
\usepackage{todonotes} 

\newcommand{\algname}[1] {{\fontfamily{cmtt}\selectfont {#1}}}

\def\plaintitle{SIGCHI Extended Abstracts Sample File: Note Initial
  Caps} 
\def\emptyauthor{}
\def\plainkeywords{Recommender systems; fairness; music recommendation; popularity bias}

\title{Unfair Exposure of Artists in Music Recommendation}

\numberofauthors{6}
\author{%
  \alignauthor{%
    \textbf{Himan Abdollahpouri}\\
    \affaddr{University of Colorado Boulder} \\
    \affaddr{Boulder, USA} \\
    \email{himan.abdollahpouri@colorado.edu} } \vfil \alignauthor{%
    \textbf{Robin Burke}\\
    \affaddr{University of Colorado Boulder}\\
    \affaddr{Boulder, USA}\\
    \email{robin.burke@colorado.edu} }
    \vfil \alignauthor{%
    \textbf{Masoud Mansoury}\\
    \affaddr{Eindhoven University of Technology}\\
    \affaddr{Eindhoven, the Netherland}\\
    \email{m.mansoury@tue.nl} }
     }

\definecolor{linkColor}{RGB}{6,125,233}
\hypersetup{%
  pdftitle={\plaintitle},
  pdfauthor={\emptyauthor},
  pdfkeywords={\plainkeywords},
  bookmarksnumbered,
  pdfstartview={FitH},
  colorlinks,
  citecolor=black,
  filecolor=black,
  linkcolor=black,
  urlcolor=linkColor,
  breaklinks=true,
}

\begin{document}

\CopyrightYear{2020}
\setcopyright{rightsretained}
\conferenceinfo{CHI'20,}{April  25--30, 2020, Honolulu, HI, USA}
\isbn{978-1-4503-6819-3/20/04}
\doi{https://doi.org/10.1145/3334480.XXXXXXX}
\copyrightinfo{\acmcopyright}

\maketitle

\RaggedRight{} 

\begin{abstract}
Fairness in machine learning has been studied by many researchers. In particular, fairness in recommender systems has been investigated to ensure the recommendations meet certain criteria with respect to certain sensitive features such as race, gender etc. However, often recommender systems are multi-stakeholder environments in which the fairness towards all stakeholders should be taken care of. 
It is well-known that the recommendation algorithms suffer from popularity bias; few popular items are over-recommended which leads to the majority of other items not getting a proportionate attention. This bias has been investigated from the perspective of the users' and how it makes the final recommendations skewed towards popular items in general. In this paper, however, we investigate the impact of popularity bias in recommendation algorithms on the provider of  the items (i.e. the entities who are behind the recommended items). Using a music dataset for our experiments, we show that, due to some biases in the algorithms, different groups of artists with varying degrees of popularity are systematically and consistently treated differently than others.
\end{abstract}




\section{Introduction}
Recommender systems have been widely used in a variety of different domains such as movies, music, online dating etc. Their goal is to help users find relevant items which are difficult or otherwise time-consuming to find in the absence of such systems.
Music streaming services such as Spotify and Pandora have become extremely popular due to their effective recommendations and helping listeners find relevant songs and artists. 

One of the limitations of the recommendation algorithms is the problem of popularity bias \cite{abdollahpouri2017controlling}: popular items are being recommended too frequently while the majority of other items do not get the deserved attention. 
This bias and methods to tackle it have been studied by many researchers but its impact on other stakeholders of the recommendations has yet to be explored. 

In this paper, we investigate the impact of popularity bias on the fairness of the recommendations from the perspective of the providers of the items. We use a sample of the LastFM music dataset created by Kowald et al. \cite{dominik2019unfairness} containing the information about users listening history to different song tracks. The providers in this case are artists since they are the ones who provided the songs. We intend to analyze how different groups of artists with different degrees of popularity are being served by these algorithms. We use several well-known recommendation algorithms including a neighborhood-based model based called User-based collaborative filtering (\algname{UserKNN}), a matrix factorization based model called non-negative matrix factorization (\algname{NMF}), a simple user-item average technique which predicts ratings based on the average rating for any given item by a user (\algname{UserItemAvg}), a random algorithm which randomly recommends items (\algname{Random}), and most-popular item recommendation, \algname{Most-pop} which recommends the same top $N$ items to everyone, for our analysis.

\section{Related work}
The problem of popularity bias and the challenges it creates for the recommender system has been well studied by other researchers \cite{anderson2006long,brynjolfsson2006niches,longtailrecsys}. Authors in the mentioned works have mainly explored the overall accuracy of the recommendations in the presence of long-tail distribution in rating data. Moreover, some other researchers have proposed algorithms that can control this bias and give more chance to long-tail items to be recommended \cite{10.1109/TKDE.2011.15,DBLP:conf/recsys/KamishimaAAS14, flairs2019}. 

In addition, the impact of this bias on users has been studied by Abdollahpouri et al. \cite{abdollahpouriRMSE2019} where authors show users with niche taste are affected the most by the popularity bias. In this work, however, we focus on the fairness of recommendations with respect to artists' expectations. That is, we want to see how popularity bias in the input data is causing the recommendations to deviate from the true expectations of different artists. 

\section{Popularity bias in recommendation}
\begin{figure}[ht]
\begin{subfigure}{.5\textwidth}
  \centering
  \includegraphics[width=.5\linewidth]{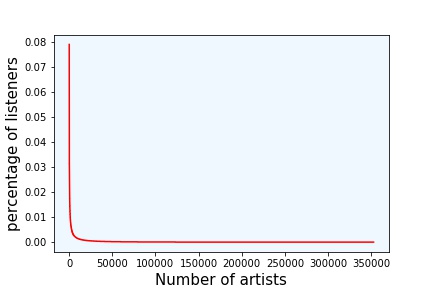}  
  \caption{Original}
  \label{fig:pop-a}
\end{subfigure}
\begin{subfigure}{.5\textwidth}
  \centering
  \includegraphics[width=.5\linewidth]{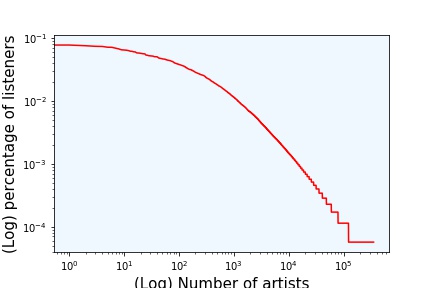}  
  \caption{Log-scale}
  \label{fig:pop-b}
\end{subfigure}
\caption{Artist popularity in LastFM data}
\label{fig:pop}
\end{figure}

Figure ~\ref{fig:pop-a} shows the distribution of artist popularity in the LastFM dataset. We can see that it has an extreme long-tail shape indicating few popular artists taking up the majority of the listening interactions. The log-scale of this plot is shown in ~\ref{fig:pop-b} for a smoother illustration. One might say these plots show that there would be no unfairness in the algorithms as users clearly are interested in certain popular artists as can be seen in the plot. However, we want to show that the algorithms are amplifying this already existing bias and it is this \textit{amplification} that we call \textit{unfair}.

\section{Fairness towards artists}
 In this paper, we call a disproportionate exposure of songs from different artists relative to what their potential listener pool could be as unfair recommendation. In other words, if songs from a certain artist could have been recommended to $X$ users but they are only recommended to $Y$ users ($|Y|<|X|$), some kind of unfair treatment exists in the recommendation algorithm. We define three different groups of artists based on their degree of popularity: 1) High-P (i.e. Mainstream), 2) Mid-P (i.e. Middle), and 3) Low-P (i.e. Niche). We used the method used in \cite{celma2008} to find the cutting points to split the artists into these groups. The number of artists fall within each group are 389, 7292, and 345,124(345K), respectively. That shows that the majority of artists have low popularity and only few artists (less than 0.01\% of the artists) fall into the \textit{Mainstream} category based on the number of times their songs are being played by the listeners (songs from these few artists take more than 30\% of the total listening counts). 
  To measure the unfairness towards each group we first define Group Average Popularity for each group which is an indication of how popular the artists in each group are on average:

\begin{equation}
    GAP(G)=\frac{\sum_{a \in G}\phi (a)}{|G|}
\end{equation}

where $\phi$ is the popularity of each artist (i.e. the number of times her songs are being played by the listeners). To measure unfairness, we then calculate the change in GAP for each group when looking at their group average popularity in data versus in recommendations. 
\begin{equation}
    \Delta GAP(G)=\frac{GAP(G)_r-GAP(G)_d}{GAP(G)_d}
\end{equation}
subscript $r$ and $d$ represent the \textit{recommendations} and \textit{training data}.

Positive values for $\Delta GAP$ show over-promotion of songs from artists belonging to a certain group while negative values indicates under-representing them.

\begin{figure}[h]
    \centering
    \includegraphics[width=2.7in]{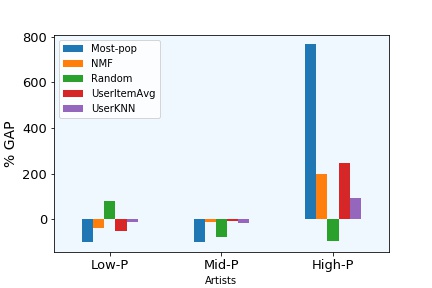}
    \caption{The $\Delta GAP$ values for three different groups of artists: Mainstream (High-P), Middle (Mid-P), and Niche (Low-P)}
    \label{fig:GAP}
\end{figure}

Figure ~\ref{fig:GAP} shows the $\Delta GAP$ for several algorithms for three different groups of artists. We can clearly see that \textit{Mainstream} group has the highest positive $\Delta GAP$ for all algorithms except for \algname{Random} indicating over-promotion of the songs from the already popular artists. \textit{Niche} and \textit{Middle} groups both have negative $\Delta GAP$ showing the suppression of songs from these groups by the algorithms. This is indeed something that needs to be addressed since the vast majority of artists fall withing these two groups. We used \algname{Random} and most popular (\algname{Most-pop}) algorithms mainly for comparison purpose to see how other algorithms perform in comparison with these two. \algname{Random} has the least bias and \algname{Most-pop} has the highest possible bias. We can see that \algname{Random} algorithm is in favor of \textit{Niche} group but hurting the other two groups with higher popularity which was expected since \algname{Random} treats all group equally while the real proportion of the interactions for these groups in the data is not equally distributed. On the other hand, the \algname{Most-pop} algorithm which only recommends top songs to everybody shows the maximum amount of unfairness for $Niche$ group and the highest bias in favor of $Mainstream$ group. Other algorithms perform somewhere in between but they all discriminate more against $Niche$ and $Middle$ while over-promoting the $Mainstream$  

\section{Conclusion and Future Work}
In this paper we argue that fairness in recommendation needs to be investigated from the perspective of all the stakeholders involved in a recommender system as these systems often are a multi-stakeholder environments \cite{abdollahpourimultistakeholder}. In this paper, we investigated the unfairness of popularity bias from the perspective of item providers: artists. We showed that the existing popularity bias in rating data can cause unfair exposure of songs from artists with different levels of popularity. Generally, in recommender systems, not much attention is given to the provider side of the products (in this case artists). However, in order for multi-sided platforms such as Spotify (listeners vs. artists), AirBnB (travellers vs hosts), eBay(buyers vs sellers) to sustain their business, it is crucial to study how these algorithms are affecting different stakeholders involved in addition to their impact on users. For future work, we will extend our analysis on more datasets, more algorithms, and we will investigate several other metrics to quantify the unfairness against item providers. 
\balance{} 

\bibliographystyle{SIGCHI-Reference-Format}
\bibliography{sample}

\end{document}